\documentclass[prb,showpacs,twocolumn,preprintnumbers,amsmath,amssymb,floatfix]{revtex4}
\usepackage[dvips]{graphicx}
\usepackage[latin1]{inputenc}
\begin{document}
\draft

\newcommand{\yc} {YbCu$_{5-x}$Au$_{x}$ }
\newcommand{\yqc} {YbCu$_{4.4}$Au$_{0.6}$ }
\newcommand{\lisi} {Li$_2$VOSiO$_4$}
\newcommand{\etal} {{\it et al.} }
\newcommand{\ie} {{\it i.e.} }
\newcommand{\aucr}{CeCu$_{5.9}$Au$_{0.1}$ }
\newcommand{\auaf}{CeCu$_{5.2}$Au$_{0.8}$ }
\newcommand{\aux}{CeCu$_{6-x}$Au$_{x}$ }
\newcommand{\ip}{${\cal A}^2$ }

\hyphenation{a-long}

\title{Magnetic field induced non-Fermi liquid to Fermi liquid crossover at the quantum critical point of YbCu$_{5-x}$Au$_{x}$}

\author{P. Carretta$^1$, R. Pasero$^2$, M. Giovannini$^2$, C. Baines$^3$}

 \affiliation{$^1$ Dipartimento di Fisica ``A.Volta", University of
Pavia, Via Bassi 6, I-27100, Pavia (Italy)}

\affiliation{$^2$ Dip. di Chimica e Chimica Industriale, University of Genova, I-16146 Genova (Italy);}
\affiliation{LAMIA-INFM-CNR, Corso Perrone 24, I-16152 Genova (Italy)}

\affiliation{$^3$ Laboratory for Muon Spin Spectroscopy, S$\mu$S
Swiss Muon Source,PSI, CH-5232 Villigen (Switzerland)}

\widetext

\begin{abstract}

The temperature (T) dependence of the muon and $^{63}$Cu nuclear spin-lattice relaxation rates $1/T_1$ in \yqc is
reported over nearly four decades. It is shown that for $T\rightarrow 0$ $1/T_1$ diverges following the behaviour
predicted by the self-consistent renormalization (SCR) theory developed by Moriya for a ferromagnetic quantum
critical point. On the other hand, the static uniform susceptibility $\chi_s$ is observed to diverge as $T^{-2/3}$
and $1/T_1T\propto \chi_s^2$, a behaviour which is not accounted by SCR theory. The application of a magnetic
field $H$ is observed to induce a crossover to a Fermi liquid behaviour and for $T\rightarrow 0$ $1/T_1$ is found
to obey the scaling law $1/T_1(H)= 1/T_1(0)[1+(\mu_BH/k_BT)^2]^{-1}$.

\end{abstract}

\pacs {76.60.Es, 71.27.+a, 75.40.Gb} \maketitle
\narrowtext

Strongly correlated electron systems with competing interactions are known to show rather rich phase diagrams,
with crossovers or phase transitions which depend on the relative magnitude of the competing energy scales. A
paridgmatic example is represented by heavy-fermion intermetallic compounds, where a quantum phase transition
between Fermi liquid (FL) and magnetic ground-states is typically observed upon varying the single-ion Kondo
coupling $J$ and the density of states at the Fermi level $D(E_F)$ \cite{HF}. The modification of these two
parameters affect both the coherence temperature $T^*$, below which the f electrons delocalize and a FL behaviour
is observed, and the transition temperature to a magnetic long-range order, which is determined by RKKY
interaction \cite{Pines}. At the quantum critical point (QCP) $T^*\rightarrow 0$, the Fermi liquid regime is never
attained and a rather peculiar behaviour of the response functions is observed down to $T\rightarrow 0$, the
so-called non-Fermi liquid (NFL) regime. The QCP can be tuned by different parameters, as the chemical
composition, the pressure and the magnetic field, which control the hybridization between f and s electron
orbitals, i.e. $J$ and $D(E_F)$ \cite{PField}. In spite of the significant experimental efforts, an overall
understanding of how the dynamical spin susceptibility behaves in the NFL regime on approaching the QCP and how it
is affected by external parameters, as the magnetic field, is still missing.

\yc is a heavy-fermion intermetallic compound which has been studied in recent years mostly with techniques of
macroscopic character, ranging from specific heat to magnetization and resistivity measurements
\cite{Yoshimura,Galli,Giova}. On the basis of these experimental results a tentative phase diagram as a function
of $x$ has been outlined. The coherence temperature $T^*$, which for $x=0$ was estimated around 5 K, vanishes
around $x\simeq 0.4$, where a quantum phase transition to a long-range magnetic order is expected
\cite{Yoshimura}. Still, it has to be pointed out that the transition temperature to the magnetically ordered
phase for $x> 0.5$ has been determined just from the change of slope in the resistivity vs. temperature
\cite{Yoshimura}, raising some doubts on the accuracy of its estimate. Moreover, a careful analysis of the
chemical and structural properties of \yc solid solutions \cite{Giovannini} have shown that homogeneous compounds
with AuBe$_5$-type structure can be grown at ambient pressure only for $x\geq 0.4$, questioning some of the
previous experimental observations.

\begin{figure}[h!]
\vspace{7cm} \includegraphics{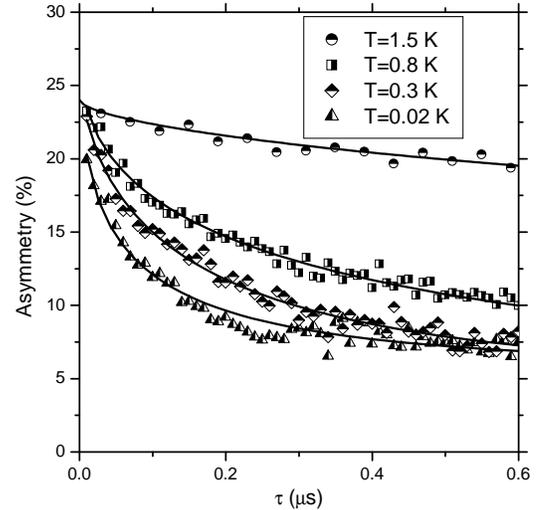}
   \caption{Decay of the muon polarization in \yqc in zero-filed at four different temperatures. The solid lines
   are the best fit according to Eq. 1. }
  \label{fig1}
\end{figure}


Here we will show, on the basis of zero and longitudinal field muon spin relaxation ($\mu$SR), nuclear quadrupole
resonance (NQR) and magnetization measurements, that for $x\simeq 0.6$ a ferromagnetic QCP is attained. Moreover,
it will be shown that the T-dependence of the muon  spin-lattice relaxation rate $1/T_1^{\mu}$  for $T\rightarrow
0$ can be suitably described within the self-consistent renormalization (SCR) theory developed by Moriya
\cite{Moriya3}. At temperatures above 1 K both $1/T_1^{\mu}$ and the copper nuclear spin-lattice relaxation rate
$1/T_1^{63}$ are observed to scale with the square of the static uniform susceptibility. Hereafter, with $1/T_1$
we shall refer both to the muon and nuclear spin-lattice relaxation rate, unless when it will be specified.
Finally, it was found that the application of a magnetic field $H$ is observed to lead to a crossover from the
critical NFL to a FL behaviour and to a significant reduction in the relaxation rate.

The experiments were performed on \yc powders grown according to the procedure reported in Ref.
\onlinecite{Giovannini}. $\mu$SR experiments were performed at PSI Swiss muon source on LTF beam line. In order to
reduce the background contamination, when the decay rate of the muon polarization $1/T_1^{\mu} \leq 1 \mu$s$^{-1}$
the data acquisition was performed in MORE mode \cite{MORE}. The decay of the muon polarization could be nicely
fit by
\begin{equation}
P_{\mu}(t)= A exp[-(t/T_1^{\mu})^{\beta}] + B \,\,\, ,
\end{equation}
with an initial asymmetry $(A+B)\simeq 24$\% , over all the
explored T range (Fig. 1). Here $B$, of the order of a few percent
is the background contribution due to the sample holder and
cryostat environment. No evidence of a transition to a
magnetically ordered phase was observed down to 20 mK, at variance
with previous findings based on resistivity measurements
\cite{Yoshimura}. As we shall see from the magnetic filed
dependence of $1/T_1^{\mu}$, the muon relaxation is dynamical,
namely  driven by spin fluctuations and not by a static field
distribution.

\begin{figure}[h!]
\vspace{8.5cm} \includegraphics{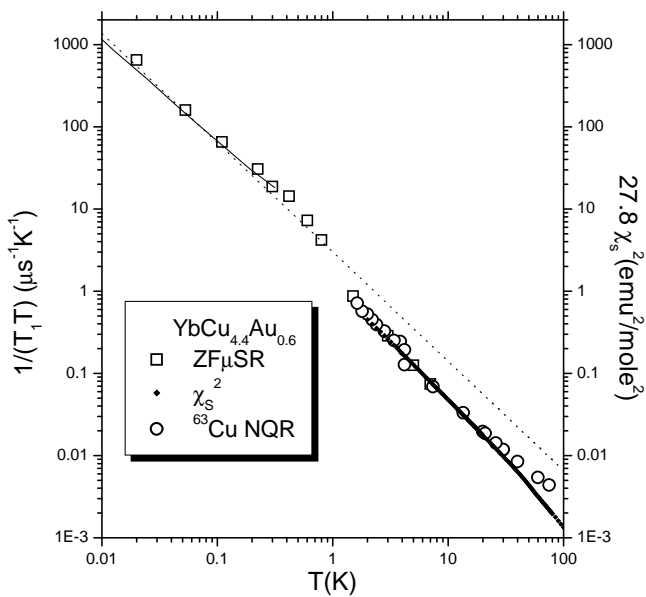}
   \caption{T-dependence of the muon and nuclear spin-lattice relaxation rates divided by temperature in \yqc , for H=0.
   The nuclear spin-lattice relaxation rate has been rescaled by a factor 236 in order to match muon $1/T_1T$ in the T-range where both
   sets of data are present. The dashed line shows the numerical results by Ishikagi and Moriya reported in Ref.\onlinecite{Moriya2}. The dotted line
   shows the power-law $T^{-4/3}$.}
  \label{fig2}
\end{figure}


$^{63}$Cu NQR $1/T_1^{63}$ measurements were performed using a
standard saturation recovery pulse sequence. The recovery law of
the nuclear magnetization after the saturating sequence could be
fit by a single exponential $y(t)= exp(-3t/T_1^{63})$ over more
than a decade. The $^{63,65}$Cu NQR spectrum, centered around 10
MHz, is broader (full width at half maximum $\simeq 2$ MHz) than
the one reported for $x=0$ \cite{NQR}, but no significant
variation of $^{63}$Cu $1/T_1$ was detected through all the
spectrum. Finally, the magnetization measurements were performed
in a magnetic field $H=100$ Gauss and the static uniform
susceptibility derived from the ratio $\chi_s= M/H$.

In Fig.2 the T-dependence of $1/T_1T$ is reported over nearly four decades. First of all one notices a trend quite
different from the one reported for $x=0$\,\,\, \cite{NQR}, where a FL ground-state is present and $1/T_1T$ gets
constant for $T\rightarrow 0$. Moreover, in Fig. 2 two other relevant aspects are evidenced: $1/T_1T$ scales with
$\chi_s^2$ and, for $T\rightarrow 0$, $1/T_1T\propto T^{-4/3}$. Accordingly, one has to expect that at low T
$\chi_s\propto T^{-2/3}$. Indeed, from Fig. 3 one notices that in the low T-range the static uniform
susceptibility $\chi_s= C/(T^{2/3} + \Theta^{2/3})$ for $x=0.6$, while this scaling is not so well obeyed for
lower or higher Au contents. Moreover, one can notice that for $x=0.6$ the Curie-Weiss temperature $\Theta$
vanishes. This indicates that if $x=0.6$ corresponds to the critical doping the quantum phase transition should
separate a FL from a ferromagnetic ground-state. This situation is somewhat analogous to the one observed in other
intermetallic compounds at the QCP, as CeCu$_{6-x}$Au$_x$ for instance \cite{Aeppli}. In this system, however, the
non-uniform susceptibility $\chi_s(q\neq 0) \propto T^{-\alpha}$ ($\alpha\simeq 0.7$ heuristically determined) and
the Curie-Weiss temperature vanishes at a wave-vector $q\neq 0$. It is worth to mention that while there is no
experimental determination of which type of order develops in \yc\, for $x$ just above 0.6, for $x=1$ an
incommensurate ordering at a wave vector $\vec Q$ far from the Brillouin zone (BZ) center was evidenced
\cite{NSDiff}. Thus, it has to be expected that $\vec Q$ approaches the BZ center as  $x\rightarrow 0.6$.

\begin{figure}[h!]
\vspace{7cm} \includegraphics{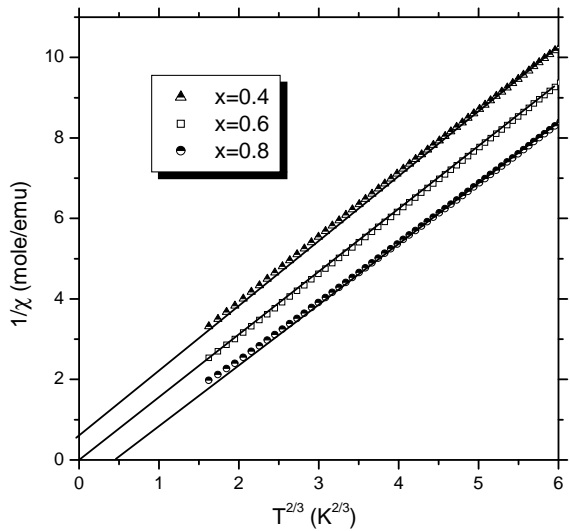}
   \caption{Inverse of the molar spin susceptibility in \yc vs. $T^{2/3}$. One notices that for $x=0.6$ a good scaling is found and that the
   Curie-Weiss temperature vanishes. }
  \label{fig3}
\end{figure}


Let us now discuss the low-energy excitations in the light of the results obtained from $1/T_1$ measurements. In
the presence of a relaxation mechanism driven by spin fluctuations one can write $1/T_1$ in terms of the imaginary
part of the dynamical spin susceptibility $\chi"(\vec q, \omega_R\rightarrow 0)$
\begin{equation}
\frac{1}{T_1}= \frac{\gamma^2 k_BT}{2N} \sum_{\vec q} |A_{\vec
q}|^2 \frac{\chi"(\vec q, \omega_R)}{\omega_R} \,\,\, ,
\end{equation}
where $\gamma$ is the gyromagnetic ratio and $|A_{\vec q}|^2$ is the form factor, giving the hyperfine coupling of
the muon (or nuclei) with the spin excitations at wave-vector $\vec q$. From Fig. 2 one notices that $^{63}$Cu and
$\mu^+$ $1/T_1$ differ by a factor 236. From the above equations, taking into account that
$(\gamma_{\mu}/\gamma_{63})^2\simeq 144$, one realizes that the hyperfine couplings of $^{63}$Cu nuclei and of the
interstitial $\mu^+$ are quite similar.

Following Ishikagi and Moriya \cite{Moriya1} it is convenient to
write the dynamical spin susceptibility in terms of two
characteristic parameters $T_0$ and $T_A$ which characterize the
width of the spin excitations spectrum in frequency and $\vec q$,
respectively. For ferromagnetic correlations one has
\cite{Moriya1,Moriya2}
\begin{equation}
\chi(q,\omega)= \frac{\pi T_0}{\alpha_Q T_A} \frac{x}{k_B2\pi T_0 x
(y + x^2) - i\omega\hbar} \,\,\,
\end{equation}
where $x=q/q_D$, with $q_D$ a Debye-like cutoff wave-vector,
$\alpha_Q$ a dimensionless interaction constant close to unity for
a strongly correlated system, and $y=N_A/2\alpha_Q k_B T_A
\chi_s$. Here the susceptibility is per spin and in $4\mu_B^2$
units and has the dimensions of the inverse of an energy, while
$T_A$ and $T_0$ are in Kelvin. From the previous expression one
can derive $\chi "(\vec q,\omega_R)/\omega_R$ by taking the limit
$\omega_R\rightarrow 0$, since $\hbar \omega_R\ll k_B T$. Then, by
integrating $\chi "(\vec q,\omega_R)/\omega_R$ in $\vec q$, over a
sphere of radius $q_D$,  one derives
\begin{equation}
\frac{1}{T_1}= \frac{\gamma^2 A^2}{2} T \frac{3\hbar }{4\pi k_B T_A
T_0} \frac{1}{\alpha_Q} \frac{1}{2y(1+y)}
\end{equation}
Now, if $T_A\gg T$  in the T-range of interest then $y\ll 1$
\cite{Moriya1,Moriya2} and one can simplify the previous
expression in the form
\begin{equation}
\frac{1}{T_1}\simeq \gamma^2 A^2  \frac{3\hbar }{8\pi}
(\frac{T}{T_0}) \frac{\chi_s}{N_A}
\end{equation}
This expression corresponds to the one derived by Ishikagi and Moriya \cite{Moriya1,Moriya2}, provided  one takes
into account that their hyperfine coupling constants are in kOe/$\mu_B$. This is the behaviour typically observed
in the presence of a magnetic ground-state \cite{MnSi}. For a three-dimensional system approaching a ferromagnetic
QCP $\chi_s$ is expected to diverge as $T^{-4/3}$ at low-temperature and, accordingly, $1/T_1\propto T^{-1/3}$,
exactly the behaviour observed in our measurements. However, it has to be pointed out that SCR theory would
predict a spin-lattice relaxation rate $1/T_1\propto T \chi_s$, at variance with the experimental findings.
Therefore, although the T-dependence of $1/T_1$ seems to agree with predictions of SCR theory we do not find a
full consistency of our experimental findings with the theoretical expectations. As we shall see in the next
paragraph, also the magnetic field dependence of $1/T_1$ can hardly be explained by SCR theory.

\begin{figure}[h!]
\vspace{9.5cm} \includegraphics{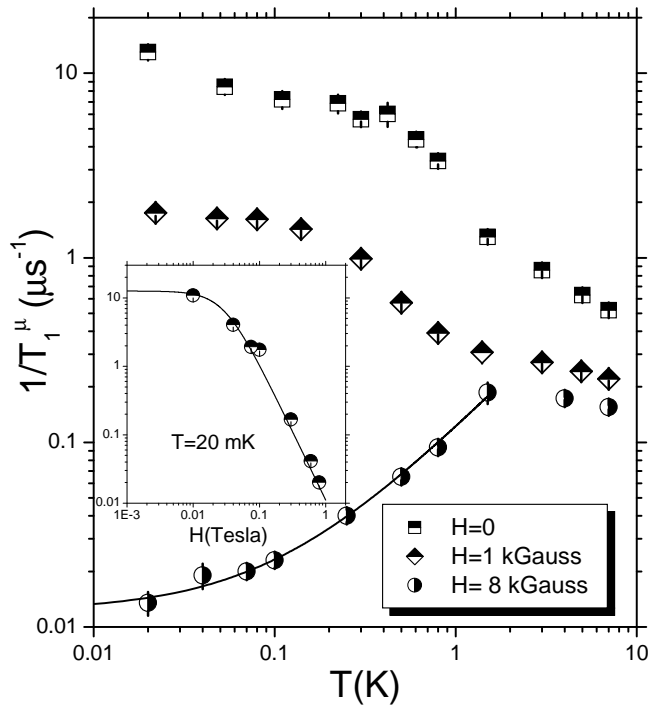}
   \caption{T-dependence of the muon  spin-lattice relaxation rate in \yqc at three different magnetic
   fields. The solid line evidences that for H=8 kGauss, at low-T,
   $1/T_1= aT + b$, where the small offset $b$ should be either
   associated with the uncertainty in the background corrections
   or to the fact that a certain angular dependence of the effect
   of the magnetic field has to be expected. In the inset the
   magnetic
   field-dependence of the muon $1/T_1$ is shown. The solid line
   shows the scaling law $1/T_1(H)= 1/T_1(0) [1+ (\mu_BH/k_BT)^2]^{-1}$, with no
adjustable parameter.}
  \label{fig4}
\end{figure}


Now we turn to the discussion of the effect of a magnetic field on the muon longitudinal relaxation rate
$1/T_1^{\mu}$. From Fig.4 one notices that the magnetic field progressively reduces $1/T_1^{\mu}$. Remarkably the
effect is significant at low T, where $1/T_1^{\mu}(H=0)$ is large, while it is reduced at high T where
$1/T_1^{\mu}(H=0)$ is low. This behaviour is the opposite of what one would expect if the relaxation had to be
associated with a static field distribution $\Delta H$. In fact, in that case one would expect a significant
reduction of the relaxation when $\gamma H\simeq 10/T_1^{\mu}(H=0)$  \cite{RDR}, at variance with the experimental
findings. Moreover, the similar behaviour of $1/T_1^{63}$ and $1/T_1^{\mu}$ suggest that the effect of the
magnetic field rather has to associated with a modification in the dynamical spin susceptibility. In fact, in
other intermetallic compounds as \aux \cite{Noi} and YbRh$_2$Si$_2$ \cite{Kita} showing an antiferromagnetic QCP a
similar effect of the magnetic field has been reported. In particular, it has been observed that the magnetic
field drives the system away from the QCP towards a FL ground-state. Here we observe that at $H= 8$ kGauss,
$1/T_1$ decreases linearly with decreasing T (Fig. 4), as expected for a FL \cite{Slichter}. Our results are
perfectly consistent with the experimental findings by Tsujii et al. \cite{Tsujii}, who observed in \yqc , for
$H\simeq 1$ Tesla, a crossover in the T-dependence of the resistivity from a $T^{3/2}$ to a $T^{2}$ power law, the
one typical of a FL.

The effect of the magnetic field cannot be explained within SCR theory, as an initial raise in $1/T_1$ and then a
decrease should be expected \cite{Moriya2}. This is not the case here, in fact, for $H=1$ kGauss $1/T_1$ increases
on cooling down to the lowest temperature, while for $H= 8$ kGauss $1/T_1$ is practically always decreasing on
cooling. In the \aux the effect of the magnetic field was accounted for by a renormalization of the temperature
scale to $T_{m}= T[1+ (\mu_BH/k_BT)^2]^{1/2}$\,\,\, \cite{Aeppli,Noi2}. Here, by taking into account that
$1/T_1\propto T\chi_s^2\propto T^{-1/3}$, one should expect that $1/T_1(H) \propto [1+ (\mu_BH/k_BT)^2]^{-1/6}$,
at variance with the experimental findings.

On the other hand, we find that the relaxation rate obeys the scaling $1/T_1(H)= 1/T_1(0) [1+
(\mu_BH/k_BT)^2]^{-1}$, with no adjustable parameter (Fig. 4). Accordingly, for $T\rightarrow 0$ $1/T_1(H)\propto
1/H^2$. Remarkably also in MnSi \cite{Gat} the muon relaxation rate was observed to decrease with $H^2$ on
approaching the transition to the magnetic ground-state. Also in that case SCR theory could not explain the field
dependence of the relaxation rate and its decrease with $H^2$ was tentatively ascribed to the progressive
quenching of the helical components of the critical fluctuations and to the increase of $q=0$ fluctuations
\cite{Gat}. This explanation, of course, cannot hold here where the critical fluctuations are at $q=0$. The
scaling law experimentally found in this work should rather take into account the progressive departure from the
ferromagnetic QCP, induced by the magnetic field, and the insurgence of the FL ground-state.

In conclusion, from the T-dependence of the muon and $^{63}$Cu
nuclear spin-lattice relaxation rates in \yqc we found that for
$T\rightarrow 0$ $1/T_1$ diverges following the behaviour
predicted by the self-consistent renormalization theory developed
by Moriya for a ferromagnetic quantum critical point and no
evidence of any phase transition could be detected down to 20 mK.
On the other hand, at low-T the  static uniform susceptibility
$\chi_s$ was observed to diverge as $T^{-2/3}$ and, accordingly,
$1/T_1T\propto \chi_s^2$, a behaviour which cannot be explained
within SCR theory. Finally, the application of a magnetic field
$H$ is observed to induce a crossover to a Fermi liquid behaviour
and for $T\rightarrow 0$ $1/T_1$ is found to obey the empirical
scaling law $1/T_1(H)= 1/T_1(0)[1+(\mu_BH/k_BT)^2]^{-1}$.
\\

The authors acknowledge useful discussions with M. Graf. The NMR
measurements in Pavia were supported by Fondazione CARIPLO2005
Research Funds.




\begin{references}

\bibitem{HF} N. Grewe and F. Steglich, in \textit{Handbook on the Physics
and Chemistry of Rare Earths}, Eds. K.A. Geschneider and L. Eyring
(North-Holland, Amsterdam, 1991), Vol. 14, p. 343
\bibitem{Pines} Y. Yang, Z. Fisk, H.-O. Lee, J.D. Thompson and D.
Pines, Nature 454, 611 (2008)
\bibitem{PField} see for instance O. Stockert, F. Huster, A.
Neubert, C. Pfeiderer, T. Pietrus, B. Will and H. v. L\"ohneysen,
Physica B 312-313, 458 (2002); K. Heuser, E.-W. Scheidt, T.
Schreiner and G.R. Stewart, Phys. Rev. B 57, R4198 (1998)
\bibitem{Yoshimura} K. Yoshimura,T. Kawabata, N. Sato, N. Tsujii,
T. Terashima, C. Terakura, G. Kido and K. Kosuge, J. Alloys Comp.
317-318, 465 (2001)
\bibitem{Galli} M. Galli, E. Bauer, St. Berger, Ch. Dusek, M.
Della Mea, H. Michor, D. Kaczorowski, E.W. Scheidt and F.
Marabelli, Physica B 312-313, 489 (2002)
\bibitem{Giova} M. Giovannini, A. Saccone, St. M\"uller, H. Michor
and E. Bauer, J. Phys.: Condens. Matter 17, S877 (2005)
\bibitem{Giovannini} M. Giovannini, R. Pasero, S. De Negri and A.
Saccone, Intermetallics 16, 399-405 (2008)
\bibitem{Moriya3} T. Moriya, \textit{Spin Fluctuations in Itinerant
Electron Magnetism}, Springer, Berlin 1985
\bibitem{MORE} R. Abela, A. Amato, C. Baines, X. Donath, R. Erne, D.C. George, D. Herlach, G. Irminger, I.D. Reid, D. Renker, G. Solt, D. Suhi, M. Werner and U. Zimmermann,
Hyperfine Interact. 120-121, 575 (1999).
\bibitem{NQR} N. Tsujii,  K. Yoshimura and K. Kosuge, Phys. Rev. B
59, 11813 (1999)
\bibitem{Aeppli} A. Schr\"oder, G. Aeppli, R. Coldea, M. Adams, O. Stockert, H.v. L\"ohneysen, E. Bucher, R. Ramazshavili and P. Coleman,
Nature 407, 351 (2000)
\bibitem{NSDiff} E. Bauer, P. Fischer, F. Marabelli, M. Ellerby, K.A. McEwen, B. Roessli and M.T. Fernandes-Dias,
Physica B 234-236, 676 (1997)
\bibitem{Moriya1} A. Ishigaki and T. Moriya, J. Phys. Soc. Jpn. 67, 3924 (1998)
\bibitem{Moriya2} A. Ishigaki and T. Moriya, J. Phys. Soc. Jpn. 65, 3402 (1996)
\bibitem{MnSi} M.Corti, F. Carbone, M. Filibian, Th. Jarlborg, A.A. Nugroho and P. Carretta, Phys. Rev.
B 75, 115111 (2007)
\bibitem{RDR} S.J. Blundell, Contemporary Physics 40, 175 (1999)
\bibitem{Noi} P. Carretta, M. Giovannini, M. Horvatic, N. Papinutto and A. Rigamonti, Phys. Rev. B 68, 220404(R)
(2003)
\bibitem{Kita} K. Ishida, K. Okamoto, Y. Kawasaki, Y. Kitaoka, O.
Trovarelli, C. Geibel and F. Steglich, Phys. Rev. Lett. 89, 107202
(2002)
\bibitem{Tsujii} N. Tsujii, T. Terashima, C. Terakura, G. Kido, T.
Kawabata, K. Yoshimura and K. Kosuge, J. Phys.: Condens. Matter
13, 3623 (2001)
\bibitem{Slichter} C. P. Slichter in \textit{Principles of Magnetic Resonance} (Springer, Berlin, 1990) 3rd Ed.
\bibitem{Noi2} P. Carretta, M. Giovannini, M.J. Graf, N.
Papinutto, A. Rigamonti and  K. Sullivan, Physica B 378-380, 84
(2006)
\bibitem{Gat} I.M. Gat-Malureanu, A. Fukaya, M.I. Larkin, A.J. Millis, P.L. Russo, A.T. Savici, Y.J. Uemura, P.P. Kyriakou, G.M. Luke, C.R. Wiebe, Y.V. Sushko,
R.H. Heffner, D.E. Mac Laughlin, D. Andreica and G.M. Kalvius,
Phys. Rev. Lett. 90, 157201 (2003)


\end{references}
\end{document}